# Tracing the Largest Seasonal Migration on Earth

Xianwen Wang[1*], Chen Liu[1], Wenli Mao[1], Zhigang Hu[1] & Li Gu[1]

School of Public Administration & Law, Dalian University of Technology, Dalian 116085, China

*Corresponding author
E-mail: xianwenwang@dlut.edu.cn
Interactive visualization of the migration: http://xianwenwang.com/research/mig/

**Abstract:** It is estimated that over 3.6 billion passengers are travelling during the Chinese Spring Festival travel season. They leave their working cities and return their hometowns to enjoy annual family time, and back to cities after the holiday. In this study, with the massive location-based data collected from millions of smartphone users, we propose a novel method to trace the migration flow and explore the migration patterns of Chinese people. From the temporal perspective, we explore the migration trend over time during a 34-days period, about half a month before and after the Spring Festival. From the spatial perspective, the migration directions and routes are estimated and quantified, and the migration flow is visualized. The spatial range of influence of developed regions could be reflected with the destinations of migration, the migration destinations and originations have obvious characteristic of geographical proximity.

**Keywords:** migration pattern; migration flow; migration network analysis; China; Chunyun

## INTRODUCTION

Every day, every moment, people are moving from one place to another. However, there is not any day neither any moment could compare with the Spring Festival (Chinese Lunar New Year) travel season, also known as Chunyun period. Like Christmas holiday in the Western nations, Chinese New Year' holiday is the most important traditional Chinese holiday for family reunions. Chinese Chunyun period, usually begins half month before the Chinese Lunar New Year's day and lasts for around 40 days, which is, for example, from January 16 to February 24 in this year. With the largest annual movement of people, hundreds of millions of people return to their towns and villages for the holidays. Chunyun period is a good chance to look into the migration of Chinese people. Since the late 1980s, an unprecedented labor migration from rural to urban areas has occurred in China(Zhao, 1999). The movement of rural Chinese to the cities has contributed a lot to the extraordinary Chinese growth(Meng, 2014).

Previous studies on migration focused on the identification of main source and destination areas(Abel *et al.*, 2014; He, 2002; Shen, 2012; Zhu *et al.*, 2010), the socio-economic determinants of migrations(Fan, 2005; Roberts, 2001; Zhao, 1999). Various migration models were employed to make analysis based on census data in almost all previous migration studies.  The establish migration models to explain migration flows (Shen, 2012; Stillwell, 2005) include the gravity model (GM)(Ravenstein, 1885; Zipf, 1946), Spatial interaction models (Congdon, 1991; Lowry, 1966; Stillwell *et al.*, 1991; Wilson, 1967), Two-stage migration modelling (Rees *et al.*, 2004), etc.

However, restricted to the availability, integrity and consistency of census data, it is almost impossible to make deep analysis about the migration, i.e., the migration routes, the migration patterns, and temporal and spatial trend.

With Location-Based Service (LBS), smartphones provide us a new way to collect people's geographical position data both from spatial and temporal perspective (Lu *et al.*, 2011; Wang *et al.*, 2008). Computational social science (Lazer *et al.*, 2009) in the era of "Big Data" is intended to process data and run simulations about human behavior at planetary scale never before possible(Conte *et al.*, 2012).

## DATA AND METHODS

On 26 January 2014, Baidu, China's largest search engine company, created an interactive heat map that displays people's travel routes in China during the Chunyun period, which is called 'Baidu Migration', as Figure 1 shows. The data of the heat map is gathered from the locations provided by 200 million smartphone users through LBS Baidu map data source.

Baidu Migration Map is updated hourly for the travel routes in the last eight hours, displays every travel route people are traveling to and leaving from. In China, most people can reach their destination in 8 hours by plane or high-speed railway. As long as people's phone is equipped to use Baidu Maps API application, then his/her long distance move could be illustrated a line in the map.

In this study, all the hourly updated data from 16 January to 18 February 2014, when the Baidu Migration data is available, is collected and processed, except for the data at 21:00 14 February, 04:00 and 05:00 17 February and 17:00 18 February, when the data is not available. In this research, the missing data is filled in with the average number of the same time point on the day before and after, when the missing data at 17:00 18 February is filled in with the average number of two neighboring point.

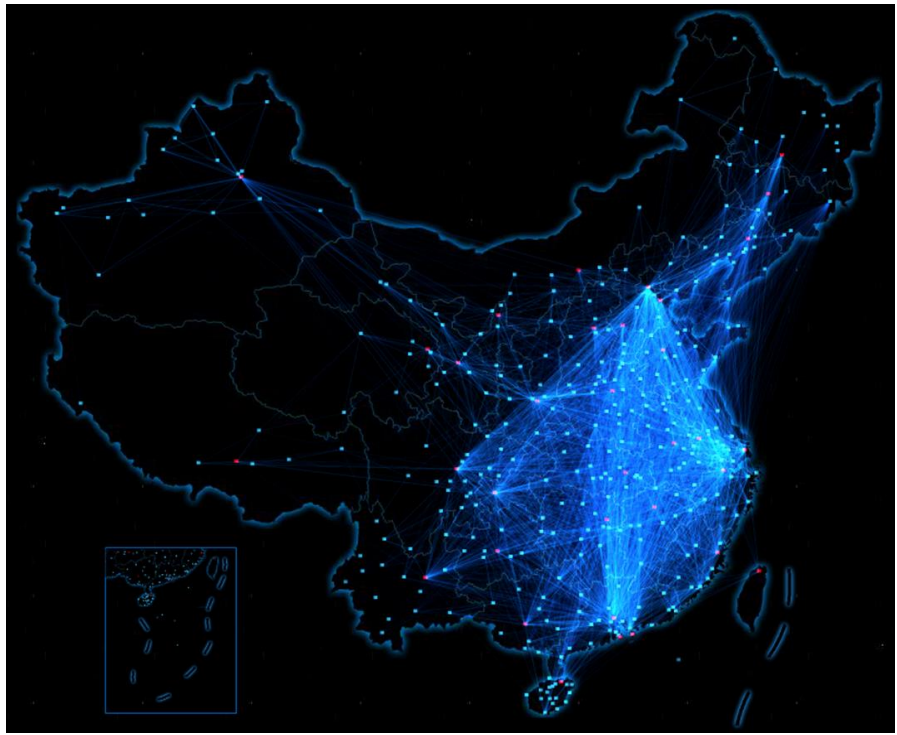

Figure 1 Baidu migration heat map

# RESULTS

**The migration trend during Chunyun period**

The hourly updated Baidu migration map provides the data during the last 8 hours; for example, at 8:00 am, the map provides the data during the period of 0:00 to 7:59 am. As Figure 2 shows, the light blue wave curve represents the hourly updated migration number during the period from 00:00 16 January to 23:00 18 February 2014, when the bold orange curve is the trend curve generated by the raw data using Hodrick–Prescott filter method, when the lambda is set to 20000.

Because the drastic fluctuations of the hourly data curve, here we focus on the trend curve. The trend curve is rather even from 16 January to 23 January. There is a significant increase on 24 January, which is Xiao Nian Day. Xiao Nian Day is the day about a week before Chinese New Year's day. Traditionally, it is the time for people to make preparations for the celebration of Chinese New Year's day. After Xiao Nian Day, the curve rises sharply and peaks on 26 January, and then the dramatic fall begins and bottoms on 30 January, when the Chinese New Year's Eve finally comes. After that the curve bounces off the two day's bottom immediately. The period of 31 January to 6 February is China's public holiday for the formal employees in government and companies. It's also the traditional time for people to visit their relatives and friends. The curve reaches a higher peak on 6 January, which is the last day of the public holiday. But for other people, e.g., the students and migrant labors, the holiday is not over. The curve starts a downtrend and reaches the second obvious bottom on 14 February, or Lantern Festival Day. It marks the end to China's half-month-long lunar New Year celebrations. After that, it starts a new peak travel. Thus you can see another obvious summit on 15 February. The migration direction is completely opposite before and after Chinese New Year's Day.

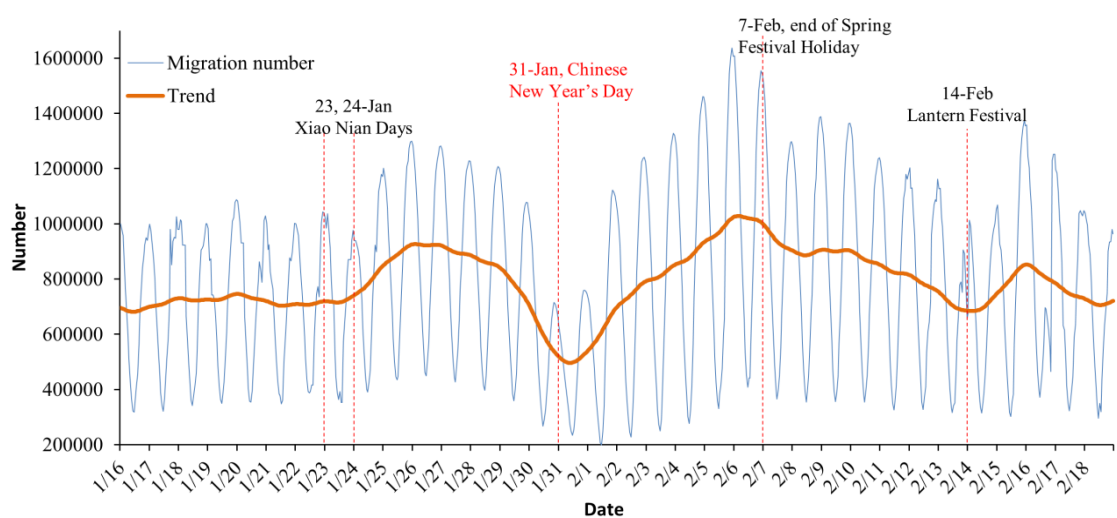

Figure 2 The Chinese migration trend during Chunyun period

**Migrating out and Migrating in**

The migration direction is completely opposite before and after Chinese New Year's Day. Before New Year's Eve, people return their hometown for reunite with their family. After that, they would go back to their work place. As Table 1 shows, Guangdong (Canton) is the top one region with most departures, followed by Jiangsu, Beijing, Hebei and Zhejiang.

Table 1 Top 5 migrating out provincial regions before Chinese New Year's Eve

| Rank | Province | Number |
|---|---|---|
| 1 | Guangdong | 2,519,061 |
| 2 | Jiangsu | 2,169,495 |
| 3 | Beijing | 1,761,509 |
| 4 | Hebei | 1,646,242 |
| 5 | Zhejiang | 1,428,408 |

Figure 3 shows the migration trend for three typical regions, including Beijing city and Shanghai city, and Guangdong province. Beijing and Shanghai is the two biggest cities in China, and Guangdong has the most factories and has been the largest province by GDP since 1989. As interpreted in Figure 2, all the three trend curves shown in Figure 3 is the result of the raw hourly data processed using Hodrick–Prescott filter method, the lambda is set to 20000.

The curve of Guangdong starts to climb as early as 18 January, which is quite different with Beijing and Shanghai. Compared with the later two cities, Guangdong has much more migrant labors, who have relative more open schedule than formal employees. The curve of Guangdong starts another sharp increase from 24 January, and peaks on 26 January, which is quite coincident with China's overall situation shown in Figure 2. For Beijing and Shanghai, which have quite similar trends, the increases between 18 and 23 January do not happen. Both two curves starts to climb on 24 January, and reach their first peak on 25 January. After that, the curves have small fluctuations at a high level during the period from 25 to 29 January. Unlike the majority of migrant labors in Guangdong, the majority in Beijing and Shanghai are formal employees, who are not easy to get the permission to leave too earlier than the national public holiday, which begins 31 January this year.

We still use 30 January as a cutoff point, for all the three curves, the left part is dramatically higher than the right half, especially for Guangdong province.

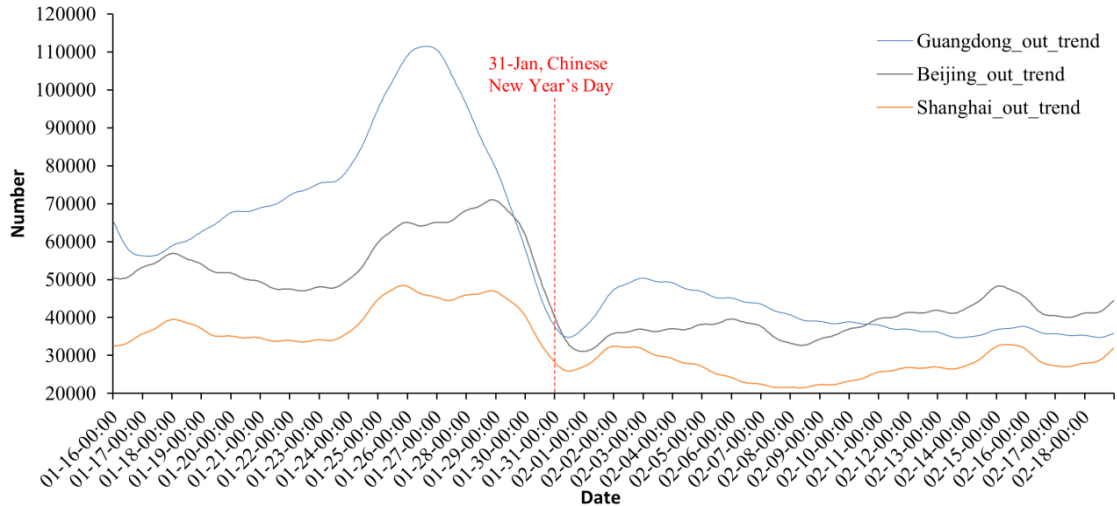

Figure 3 The flowing out trend of three regions

Taking Guangdong as example, we consider both the migrating out and migrating in trend together, as Figure 4 shows. Given the migrating in data positive value (the blue curve) and the migrating out data negative value (the orange curve), the net flow data is calculated by the migrating in data minus the migrating out data (the gray-tinted area). Before 31 January, the number of daily migrating out population is much more than the migrating in population, so the gray-tinted area under the x axis indicates the net outflow. However, after 3 February, this flow direction is going into reverse. The number of people migrating in Guangdong to go back to their working factories is far exceed the number of people migrating out, when the gray-tinted area above the x axis indicates the net inflow.

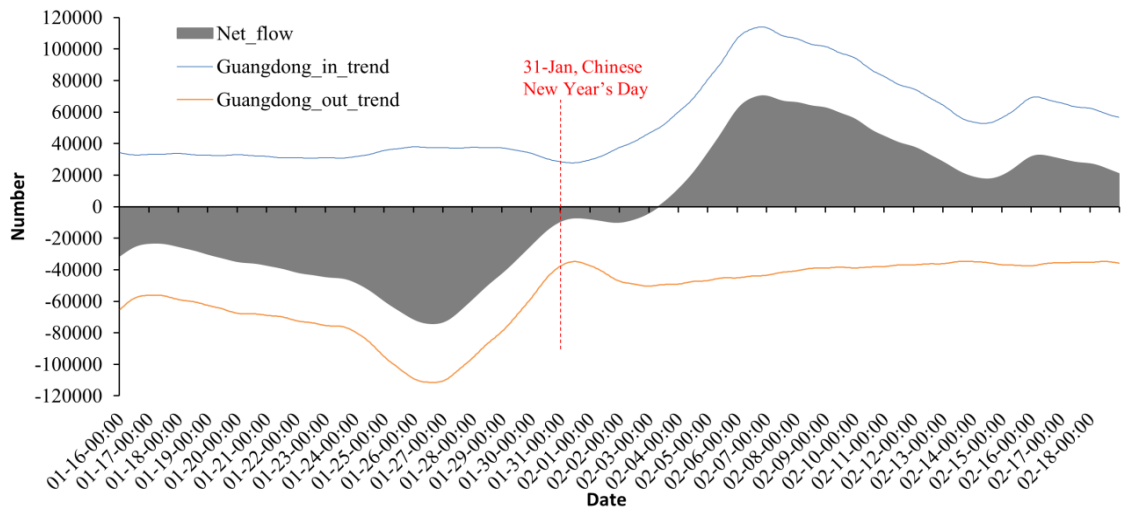

Figure 4 The migrating out, migrating in and net flow of Guangdong province

If we consider another type of region with many migrant labors originated, e.g., Guangxi Zhuang Autonomous Region (Guangxi Province), the migration trend curves are shown in Figure 5. The shape of the two curves and the gray-tinted area are quite opposite with the situation in Guangdong province shown in Figure 4. Before Chinese

New Year's Day, the migration is dominated by the migrating in streams. After that, the population flow changes to the opposite direction.

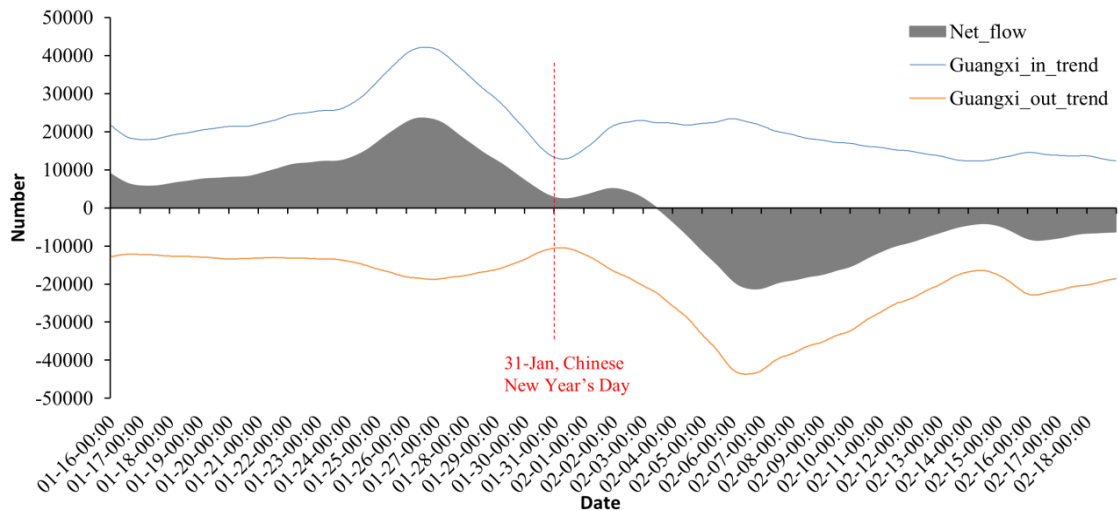

Figure 5 The migrating out, migrating in and net flow of Guangxi province

More specifically, how are the bidirectional migration flows between two regions? For example, the bidirectional migration trend between Guangdong and Guangxi. As shown in Figure 6, the blue curve indicates the migration direction of Guangdong to Guangxi, when the orange curve illustrates the migration flow from Guangxi to Guangdong. Each curve has a mountain but in different time period, the mountain of the blue curve is located at the left half part, when the orange mountain is on the right part of the curve.

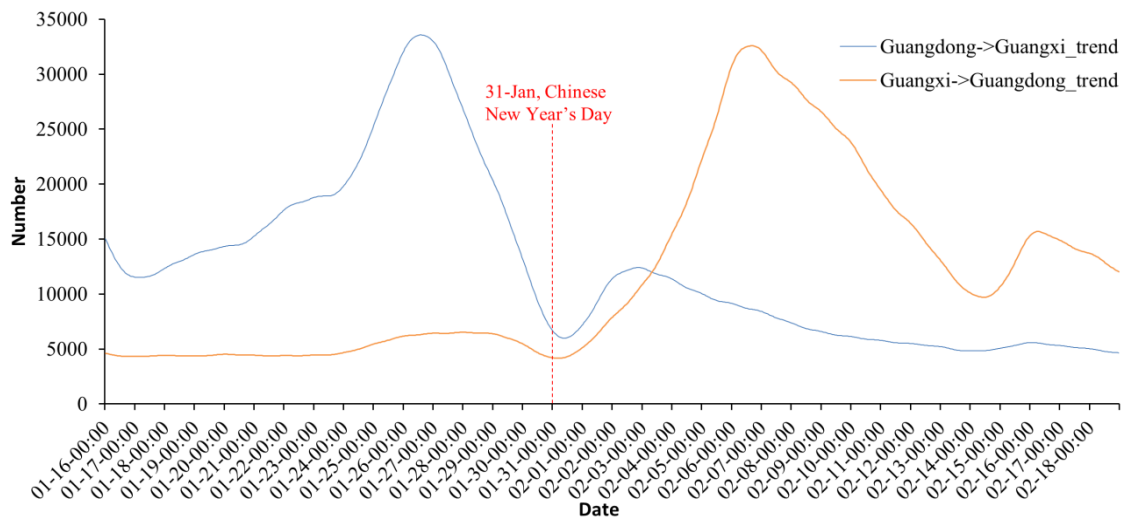

Figure 6 The bidirectional migration trend between Guangdong province and Guangxi province

**Migration network analysis**

The migration flow among 34 provincial regions in China could be described as a matrix. Table 2 shows the migration matrix among 8 regions at the time point of 8:00 AM on January 20, 2014. Value in each cell in the matrix represents the migration number from the above region to the left region during the previous 8 hours. For example, the number "1984" under Tianjin means that there are 1984 people moving from Tianjin to Beijing.

Table 2 Migration matrix among 8 regions at 8:00 AM on January 20, 2014

|  | Beijing | Tianjin | Hebei | Zhejiang | Shanghai | Jiangsu | Guangdong | Guangxi |
|---|---|---|---|---|---|---|---|---|
| Beijing | NA | 1984 | 7781 | 513 | 317 | 501 | 771 | 127 |
| Tianjin | 1601 | NA | 3946 | 208 | 70 | 159 | 82 | 10 |
| Hebei | 11650 | 4245 | NA | 117 | 62 | 567 | 419 | 56 |
| Zhejiang | 391 | 143 | 66 | NA | 3277 | 3641 | 857 | 48 |
| Shanghai | 209 | 54 | 33 | 3796 | NA | 6545 | 535 | 68 |
| Jiangsu | 443 | 261 | 518 | 4656 | 7121 | NA | 868 | 123 |
| Guangdong | 615 | 108 | 491 | 829 | 446 | 890 | NA | 3678 |
| Guangxi | 91 | 16 | 66 | 86 | 48 | 136 | 16716 | NA |

If we want to take a long period into account and get the big picture of the whole Chunyun period, the hourly updated data need to be processed specifically, because the data at each hourly point includes the total number of the past 8 hours and we can't simply sum the hourly data. In our study, the processing of raw data could be described as follows, taking the time span from January 20 to January 29 as the whole period before Chinese New Year's Day (Before-New-Year-Day-Period), and the time span from February 7 to February 18 as the whole period after Chinese New Year's Day (After-New-Year-Day-Period). Data from January 30 to February 6 is not included, because many people would travel to visit relatives and friends during these days and may result in bias to the data. For each period of the two long periods before and after Chinese New Year's Day, the hourly updated data is selected randomly to be included in our dataset. Finally, data at 30 random time points are selected into the dataset of the Before-New-Year-Period, when the After-New-Year-Period consists of data at 34 random time points. Then, the two matrices about where and how many people to and from during two periods are processed in our dataset (see supporting information).

After that, matrix is visualized as a circle divided into arcs; each arc represents a provincial region in China, as Figure 7 shows. The length of arc correlates with the number of migrating out of the corresponding region. The arcs are connected with links that represent the flow of people between two regions, when the thickness of the links encodes the relative frequency of migration. The color of each link is determined by the region that has the relative more migrants. In order to keep the graphic clean, we set a threshold of 10,000 people, which means that only those links with migration more than 10,000 people are kept, other links with less data are excluded.

The visualization is built with d3.js, and Figure 7 is an interactive graph when opening it in browser. The interactive version of Figure 7 is available at the corresponding author's website (http://xianwenwang.com/research/mig/). Select a province by moussing over the graphics to see the number of people moving between the selected province and another; it is very clear to explore the migrations between any two regions. Figure 7(a) indicates the migration network during the period before 2014 Chinese New Year's Day, which is from 20 January to 29 January specifically, the period for migrant labors to return their home villages. Figure 7(b) represents the migration network after Chinese New Year's Day, here we choose the period from 07 to 18 February correspondingly, when the migrant labors go back to work. In the interactive version, when moussing over the edge of Guangdong, the migration from and to Guangdong could be highlighted, as Figure 7(c) and Figure 7(d) show.

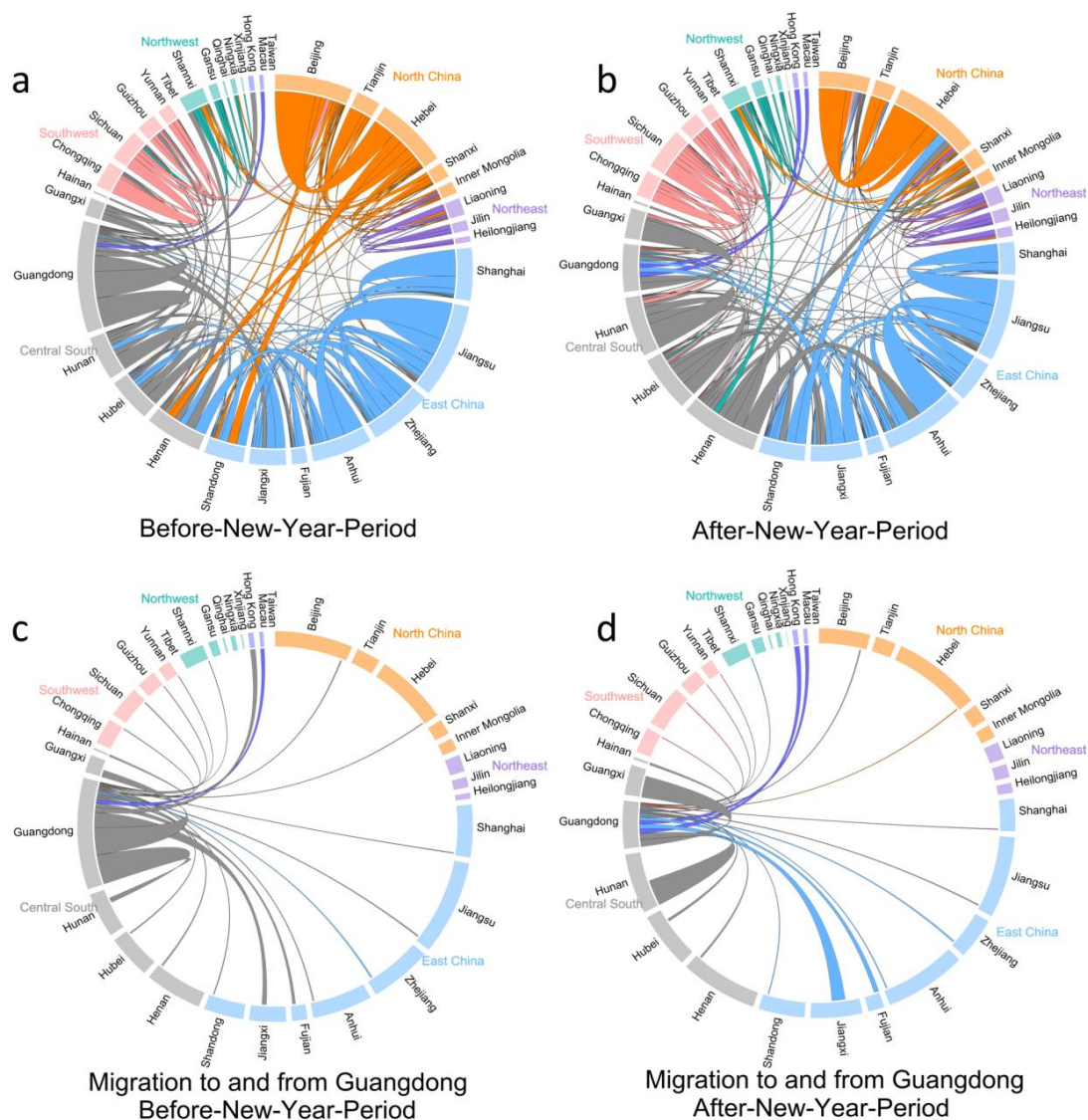

Figure 7 Chord Diagram of migration network before and after Chinese New Year's Day (a-d). The visualization is built with d3.js Explore interactive versions of the maps at http://xianwenwang.com/research/mig/.

Here we still take Guangdong province, Beijing city and Shanghai city as examples. As Figure 8 shows, the green dots indicate the destination of people migrating out from Guangdong province, the red dots for Shanghai, and the blue dots for Beijing. The migration source and destination regions have obvious characteristic of geographical proximity, as Figure 8 shows.

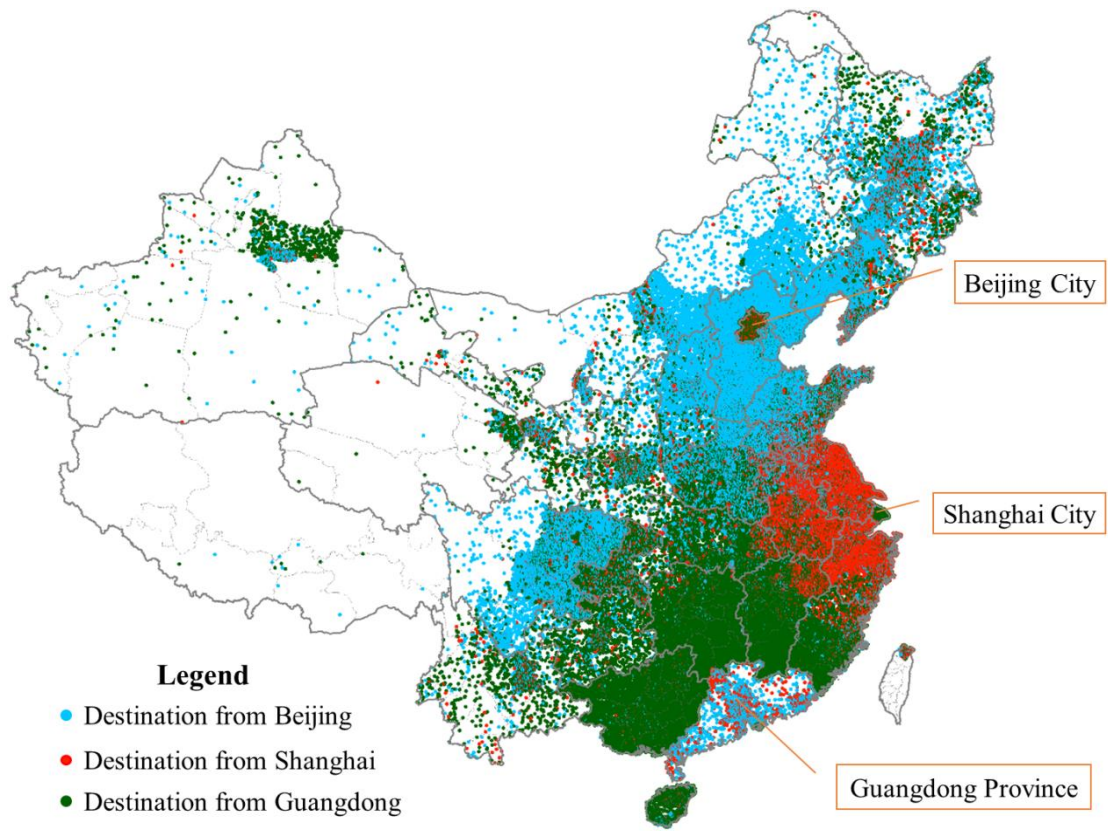

Figure 8 The destinations from Beijing, Shanghai and Guangdong before Chinese New Year's Day

## DISCUSSION

During the past 30 years, trips made by Chinese people during China's Spring Festival holiday have increased from 1 million to 3.6 billion. It is a very interesting thing to look into the temporal and spatial trends of the migrant in detail. In this research, we use the millions of LBS data during China's Spring Festival travel season to capture the largest global seasonal migration on earth for the first time. From the temporal perspective, the migration population has remarkable regularity. Take Chinese New Year's Day as the most crucial cut-off point, the travel peak begins approximately one week before the cut-off point and last till the very day before Chinese New Year's Eve. After the cut-off point, the migration population starts a new increase and reaches another peak just before the end of national public holidays of Spring Festival. After that, the number decreases slowly but still keeps at a high level till the Chunyun period ends. From the spatial perspective, we explore the travelling routes and migration destinations. The migration direction between regions before and after Chinese New Year's Day is just opposite. The spatial range of influence of developed regions could be reflected with the destinations of migration, our results reveal that even for the most developed regions in China, the migration destinations and originations have obvious characteristic of geographical proximity.

With the booming of the market of smart devices, especially smartphones in China, more and more data is generate from users. Big data has received great attention from every walk of life in China. Chinese big IT companies, including Baidu, Tencent,

Alibaba and Sina, have accumulated massive volume of data. However, what does the data be used for and how to make analysis of the data still remain a big problem, which need the academia, industry and government to work together to solve.

**ACKNOWLEDGEMENTS**

This work was supported by the project of "National Natural Science Foundation of China" (61301227).

**Note**

We have been working on this research and since January, 2014. We submitted the manuscript to Science on April 4 2014, and rejected by Science after the in-depth review.